\documentclass[showpacs,preprintnumbers,amsmath,amssymb]{revtex4}
\usepackage{graphicx}% Include figure files
\usepackage{dcolumn}% Align table columns on decimal point
\usepackage{bm}% bold math

\def\be{\begin{equation}}
\def\ee{\end{equation}}
\def\bea{\begin{eqnarray}}
\def\eea{\end{eqnarray}}
\def\NO{\nonumber}
\def\gev{\mathrm{~GeV}}

\begin{document}

%\preprint{APS/123-QED}

\title{$h_c$ Production at Hadron Colliders}% Force line breaks with \\

%\author{}%
%\affiliation{
%}%

\author{Jian-Xiong Wang$^{1,2}$}
\author{Hong-Fei Zhang$^{3,1,2}$}
\email{hfzhang@ihep.ac.cn}
\affiliation{
$^{1}$ Institute of High Energy Physics, Chinese Academy Sciences, P.O. Box 918(4), Beijing, 100049, China.\\
$^{2}$ Theoretical Physics Center for Science Facilities, CAS, Beijing, 100049, China.\\
$^{3}$ Department of Physics, School of Biomedical Engineering, Third Military Medical University, Chongqing 400038, China.
}%
\date{\today}% It is always \today, today,

\begin{abstract}
In this paper, we present the study of the hadroproduction rate of $h_c$ at next-to-leading order in $\alpha_s$ under the nonrelativisitic QCD (NRQCD) factorization framework,
using color-octet long-distance matrix elements obtained from a global fit of experimental measurements on $\chi_c$ yield and the ratio $d\sigma(\chi_{c2})/d\sigma(\chi_{c1})$
from the CDF, LHCb, CMS, and ATLAS Collaborations.
This paper considers the problem of NRQCD scale dependence for the first time,
and find that, for some experimental conditions, the choice of this scale can significantly affect the final results,
which indicates that, for these conditions, theoretical evaluation up to next-to-leading order cannot provide sufficiently precise predictions.
We also present a brief analysis on NRQCD scale dependence problem,
and provide a criterion to determine in which case next-to-leading order prediction would be ruined by the scale dependence.
\end{abstract}

\pacs{12.38.Bx, 12.39.St, 13.85.Ni, 14.40.Pq}% PACS, the Physics and Astronomy
%\keywords{Suggested keywords}%Use showqkeys class option if keyword
\maketitle
\section{introdution}
In the last ten years, many experimental measurements for P-wave quarkonia
$h_c, h_b (1^{+-})$($^1P_1$ charmonium and bottomonium states) have been achieved.  
The related branching ratios~\cite{Calderini:2004pd, Andreotti:2005vu, Fang:2006bz, Ablikim:2010rc, Lees:2011zp, Ge:2011kq, Ablikim:2012ur},
the masses of these quarkonia~\cite{Rosner:2005ry, Rubin:2005px, Dobbs:2008ec, CLEO:2011aa, Adachi:2011ji},
as well as the cross sections for $h_c (h_b)$ production via $e^{+}e^{-}$ annihilation at the CLEO-c~\cite{CLEO:2011aa} and B-factories~\cite{Adachi:2011ji} are measured.
By contrast, only leading order (LO) results have been given for $h_c$ ($h_b$) productions.
The calculations of $h_c$ hadroproduction at the Tevatron~\cite{Sridhar:1996vd} and LHC~\cite{Sridhar:2008sc, Qiao:2009zg} predicted a significant yield.
Photoproduction of $h_c$ was investigated in Ref.~\cite{Fleming:1998md}
by using a color-octet (CO) long-distance matrix element (LDME) extracted from the decay $B\rightarrow\chi_{cJ}+X$;
the results indicated a significant cross section at the DESY HERA.
$h_c$ production via $e^{+}e^{-}$ annihilation was investigated in two recent papers~\cite{Wang:2012tz, Jia:2012qx},
the former of which presented the results for $h_b$ as well.

The lack of works on $h_c$ reveals the overlook of the importance of this meson.
First of all, the hadroproduction rate of $h_c$ is a good test of nonrelativistic QCD (NRQCD)~\cite{Bodwin:1994jh}, since, for one thing,
the prediction based on color-singlet model (CSM) is far below the NRQCD result;
one is easy to favor one over the other by comparing them to experimental measurement.
And for another thing, the cross sections depend on only one nonperturbative parameter, which is not like the case of $J/\psi$,
where the difficulty of the precise determination of the three CO LDMEs causes ambiguity
~\cite{Gong:2012ug, Chao:2012iv, Ma:2010yw, Butenschoen:2012px}.
Moreover, since the branching ratio $B(h_c\rightarrow \eta_c+\gamma)$ is as large as $50\%$~\cite{Ablikim:2012ur, Ge:2011kq},
precise prediction of prompt $\eta_c$ production rate requires the evaluation of $h_c$ production rate.
One should notice that, $\eta_c$ inelastic production in $ep$ collisions is another one of the best processes
for testing NRQCD other than the $J/\psi$ hadroproduction~\cite{Hao:1999kq, Hao:2000ci}.
Finally, NRQCD scaling rule requires the CO LDME for $h_c$ should be of the same magnitude as that for $\chi_{c1}$.
This rule does not only provide an opportunity to investigate $h_c$ in NRQCD framework despite the lack of experimental measurement,
but also provides an opportunity to test the corresponding velocity scaling rule.

Since next-to-leading order (NLO) corrections noticeably change the behavior of the transverse momentum ($p_t$)
distribution of the production rate of P-wave quarkonia through color-singlet (CS) channel~\cite{Ma:2010vd, Gong:2012ug}.
LO preditions cannot reach a sufficient precision for the evaluation of $h_c$ production rate at hadron colliders.
As a result, calculation at NLO is necessary and important.

This paper is devoted to the study of the $p_t$ distribution of $h_c$ hadroproduction at NLO of QCD coupling constant $\alpha_s$.
Since there is no adequate experimental data to extract the CO LDME for $h_c$ production ($\langle O^{h_c}(^1S_0^{[8]})\rangle$) directly,
thanks to the spin symmetry at LO of NRQCD expansion, we simply estimate this LDME as $\langle O^{\chi_{c1}}(^3S_1^{[8]})\rangle$,
of which both LO and NLO values have already been given in sevaral papers~\cite{Cho:1995vh, Cho:1995ce, Ma:2010vd, Gong:2012ug};
the latter three employed experimental data of CDF~\cite{Abe:1997yz},
while the first one added the LHCb data~\cite{LHCb:2012af} and carried out a global fit.
Ref.~\cite{Jia:2014jfa} updated the fit by including all the experimental measurements available for $\chi_c$ production.
It employs $\chi_c$ production rate in Ref.~\cite{Aaij:2011jh, LHCb:2012af, ATLAS:2014ala},
and the cross section ratio $\sigma(\chi_{c2})/\sigma(\chi_{c1})$ in Ref.~\cite{Abulencia:2007bra, Aaij:2013dja, Chatrchyan:2012ub},
and presents a detailed analysis on the fit procedure and the validity of the LDME.
Here, we will employ the values of the LDME provided by Ref.~\cite{Jia:2014jfa}.

Another thing we want to address before the discussions is that,
CSM cannot provide NLO corrections to $\chi_c$ hadroproduction, since the infrared (IR) divergence does not cancel.
In NRQCD framework, this part of the IR divergence is canceled by including the NLO corrections to the CO LDME.
To renormalize the ultraviolet (UV) divergence coming from the correction to the LDME, one should introduce a new scale,
which here is called the NRQCD scale (denoted as $\mu_\Lambda$).
Before our work, it was believed that, $\mu_\Lambda$ dependence can be absorbed into the CO LDME,
and the final result is almost invariant when this scale varies~\cite{Ma:2010vd}.
The default choice of $\mu_\Lambda$ is the mass of the heavy quark.
However, our investigation shows that, for some experimental conditions,
the choice of this scale can significantly affect the final results.
Therefore, we employ the values of the CO LDME obtained at different $\mu_\Lambda$ choices to see how much the results depend on this scale.
As is know, an all-order (in $\alpha_s$) calculation will eliminate all the free scales from the expression of a physical quantity.
That is to say, if the NRQCD scale dependence of the final results at NLO is severe,
the calculation up to this order is not sufficient to make a precise prediction.

The rest of the paper is organized as follows.
Section II provides a brief review of the formulas for calculating $h_c$ hadroproduction under NRQCD framework.
In Section III, we present the numerical results and show how much the results depend on $\mu_\Lambda$.
Section IV consists of a brief analysis of the NRQCD scale dependence problem
and provide a criterion to determine in which case NLO prediction fails.

\section{$h_c$ production in NRQCD framework}

In NRQCD framework,
differential cross sections for $h_c$ hadroproduction can be expressed as
\bea
d\sigma(pp\rightarrow h_{c}+X)&=&df(pp\rightarrow c\bar{c}(n)+X)\langle O^{h_c}(n)\rangle \label{eqn:s} \\
&=&\sum_{i,j,n}\int dx_{1}dx_{2}G(x_{1},i)G(x_{2},j)
d\hat{f}(i+j\rightarrow c\bar{c}(n)+X)\langle O^{h_c}(n)\rangle \NO ,
\eea
where $p$ denotes either a proton or an antiproton,
the indices $i, j$ run over all partonic species,
n denotes a definite $c\bar{c}$ state of certain color and angular momentum,
$G(x,i)$ is the parton-distribution-function (PDF) with $x$ being the ratio of the momentum of parton $i$ to that of the proton or antiproton,
$\hat{f}$ represents the parton-level short-distance coefficient,
which can be evaluated perturbatively in $\alpha_s$ and $v$ (the relative velocity of quark and antiquark in quarkonium).
Since our calculation is up to LO in $v$, only two channels $^1P_1^{[1]}$ and $^1S_0^{[8]}$ are involved.

The LO partonic processes are listed as
\be
g+g\rightarrow c\bar{c}(^{1}P^{\left[1\right]}_{1})+g , \label{eqn:lcs}
\ee
and
\bea
&&g+g\rightarrow c\bar{c}(^{1}S^{\left[8\right]}_{0})+g, \NO \\
&&g+q(\bar{q})\rightarrow c\bar{c}(^{1}S^{\left[8\right]}_{0})+q(\bar{q}), \label{eqn:lco} \\
&&q+\bar{q}\rightarrow c\bar{c}(^{1}S^{\left[8\right]}_{0})+g. \NO
\eea
The procese in Eq.(\ref{eqn:lcs}) generates $^1P_1^{[1]}$ state, while the three processes in Eq.(\ref{eqn:lco}) generate $^1S_0^{[8]}$ state.
Both of the two states transit into $h_c$ through long-distance processes which can not be evaluated perturbatively, and the transition rates are described by LDMEs.
The LDME of $^1P_1^{[1]}$ can be expressed in terms of radial derivative of the wave function of quarkonium at the origin~\cite{Bodwin:1994jh}:
\be
\langle O^{h_c}(^1P_1^{[1]})\rangle=\frac{27}{2\pi}|R_{h_c}'(0)|^2,
\ee
while $\langle O^{h_c}(^1S_0^{[8]})\rangle$ is obtained from the fit of experimental measurements.
To evaluate the short-distance coefficients, we notice they are independent of the long-distance asymptotic states, and replace $h_c$ in Eq.(\ref{eqn:s}) by an on-shell heavy quark pair state with definite quantum numbers,
the following equations can be obtained,
\bea
d\sigma(m)=\sum_{n}df_n\langle O^{m}(n)\rangle, \label{eqn:sig}
\eea
where, and through out the rest of this paper,
we denote the cross sections $\sigma(pp\rightarrow n+X)$ and short-distance coefficients $f(pp\rightarrow n+X)$ in abbreviation as $\sigma(n)$ and $f_n$, respectively.
$d\sigma(m)$ can be evaluated perturbatively in $\alpha_s$ by reading the amplitudes from Feynman diagrams.
For $^1P_1^{[1]}$ state, up to leading order in $v$, only two states, $^1P_1^{[1]}$ and $^1S_0^{[8]}$, are involved, thus, we have
\be
d\sigma(^{1}P^{[1]}_{1})=df_{^{1}P^{[1]}_{1}}\langle O^{^1P_1^{[1]}}(^{1}P_{1}^{[1]})\rangle
+df_{^{1}S^{[8]}_{0}}\langle O^{^1P_1^{[1]}}(^{1}S_{0}^{[8]})\rangle .\label{eqn:fac}
\ee

We should keep in mind that, Eq.(\ref{eqn:fac}) is to extract CS short-distance coefficient, which is expanded in $\alpha_s$.
As a result, the quantities $\langle O^{^1P_1^{[1]}}(^{1}P_{1}^{[1]})\rangle$ and $\langle O^{^1P_1^{[1]}}(^{1}S_{0}^{[8]})\rangle$
should also be evaluated perturbatively, and the value of $\alpha_s$ in them should be in accordance with in the short-distance coefficients.
The left and right side of Eq.(\ref{eqn:fac}) should keep those terms up to the same order in perturbative expansion.

At LO in $\alpha_s$, $\langle O^{m}(n)\rangle$ vanishes unless $m=n$,
as a result, the short-distance coefficients can be simply expressed in terms of integrated squared amplitudes.

Up to NLO in $\alpha_s$, the calculation of $^1S_0^{[8]}$ state production has been described in detail in our previous paper~\cite{Gong:2008ft},
where eleven processes are involved,
\bea
&&V: g+g\rightarrow c\bar{c}(^1S_0^{[8]})+g, \NO \\
&&V: g+q(\bar{q})\rightarrow c\bar{c}(^1S_0^{[8]})+q(\bar{q}), \NO \\
&&V: q+\bar{q}\rightarrow c\bar{c}(^1S_0^{[8]})+g, \NO \\
&&g+g\rightarrow c\bar{c}(^1S_0^{[8]})+g+g, \NO \\
&&g+g\rightarrow c\bar{c}(^1S_0^{[8]})+q+\bar{q}, \NO \\
&&g+q(\bar{q})\rightarrow c\bar{c}(^1S_0^{[8]})+g+q(\bar{q}), \NO \\
&&q+\bar{q}\rightarrow c\bar{c}(^1S_0^{[8]})+g+g, \NO \\
&&q+\bar{q}\rightarrow c\bar{c}(^1S_0^{[8]})+q+\bar{q}, \NO \\
&&q+\bar{q}\rightarrow c\bar{c}(^1S_0^{[8]})+q'+\bar{q}', \NO \\
&&q+q(\bar{q}+\bar{q})\rightarrow c\bar{c}(^1S_0^{[8]})+q+q(\bar{q}+\bar{q}), \NO \\
&&q(\bar{q})+q'(\bar{q}')\rightarrow c\bar{c}(^1S_0^{[8]})+q(\bar{q})+q'(\bar{q}'), \NO
\eea
where lable V means one-loop virtual corrections to the process on the right side of it,
and $q$ and $q'$ denote quarks with different valence.
The readers can refer to Ref.~\cite{Gong:2008ft} to find out the detail of the calculation.

Here we focus on $^1P_1^{[1]}$ state production at NLO.
The processes involved are as follows,
\bea
&&V: g+g\rightarrow c\bar{c}(^{1}P^{\left[1\right]}_{1})+g , \NO \\
&&g+g\rightarrow c\bar{c}(^{1}P^{\left[1\right]}_{1})+gg , \NO \\
&&g+g\rightarrow c\bar{c}(^{1}P^{\left[1\right]}_{1})+q\bar{q}, \label{eqn:ncs} \\
&&g+q(\bar{q})\rightarrow c\bar{c}(^{1}P^{\left[1\right]}_{1})+gq(\bar{q}), \NO \\
&&q+\bar{q}\rightarrow c\bar{c}(^{1}P^{\left[1\right]}_{1})+gg. \NO
\eea
Summing all the processes, one finds the IR divergences do not cancel,
which can be figured out in NRQCD framework by subtracting the divergence coming out of $\langle O^{^1P_1^{[1]}}(^1S_0^{[8]})\rangle$.
Up to the order maintained in our calculation,
the transition rate of $c\bar{c}$ state $^1S_0^{[8]}$ into $^1P_1^{[1]}$ can be calculated in dimensional regularization and $\overline{MS}$ renormalization scheme as
\be
\langle O^{^1P_1^{[1]}}(^{1}S_{0}^{[8]})\rangle^{NLO}=-\frac{\alpha_s}{3\pi m_{c}^2}u^{c}_{\epsilon}\frac{N_{c}^{2}-1}{N_{c}^{2}}
\langle O^{^1P_1^{[1]}}(^{1}P_{1}^{[1]})\rangle^{LO} , \label{eqn:LDM}
\ee
where $N_c$ is 3 for $SU(3)$ gauge field and $u^c_\epsilon$ is defined as
\be
u^{c}_{\epsilon}=\frac{1}{\epsilon_{IR}}-\gamma_{E}-\frac{1}{3}+ln(\frac{4\pi\mu^2}{\mu^2_{\Lambda}}) \label{eqn:uc} ,
\ee
with $\gamma_E$ being Euler's constant.
$\mu_\Lambda$ is a scale rising from the renormalization of the LDME,
and $\mu$ is the scale to complement the dimension.
The detail of the calculation can be found in Ref.~\cite{Petrelli:1997ge}.

\begin{figure}
\center{
\includegraphics*[scale=0.55]{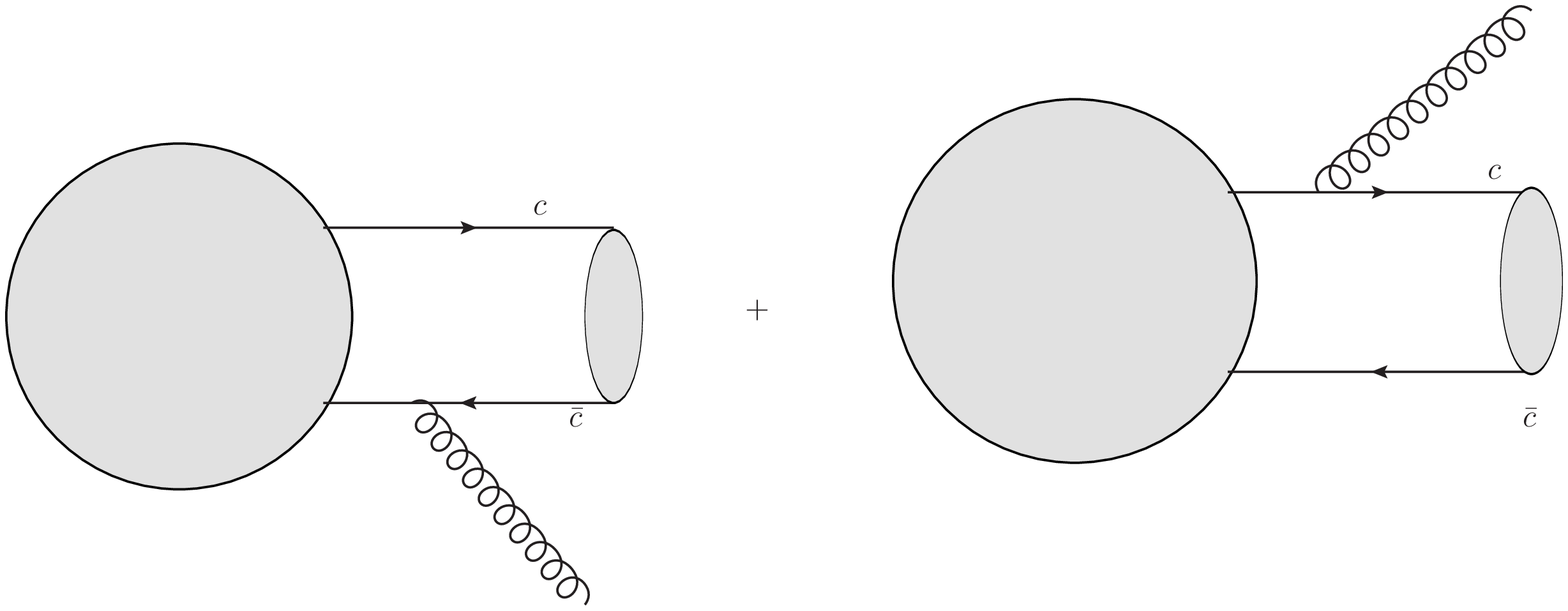} \NO \\% Here is how to import EPS art
\caption{\label{fig:gc} Typical diagrams where the soft gluon connects to the quarkonium,
the square of which corresponds to $d\sigma^{S_2}$.}
\includegraphics*[scale=0.55]{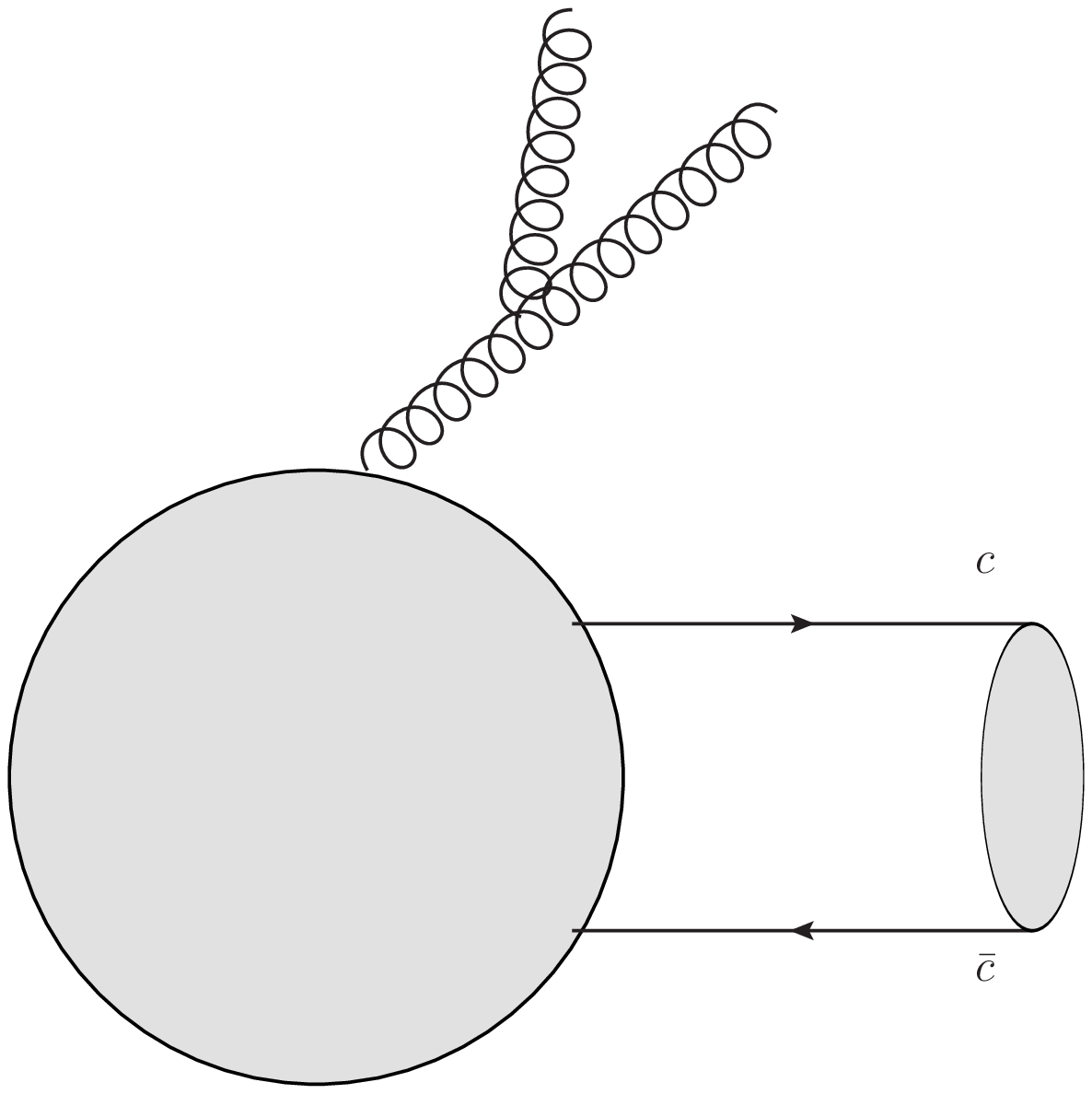}% Here is how to import EPS art
\caption {\label{fig:gg} A typical diagram where the soft gluon connects to an external particle other than the quarkonium.
}}
\end{figure}

The divergence in the LDME $\langle O^{^1P_1^{[1]}}(^{1}S_{0}^{[8]})\rangle$ will cancel that in $\sigma({^1P_1^{[1]}})$,
which can be isolated by using the two-cutoff phase space slicing method~\cite{Harris:2001sx} as
\bea
d\sigma(^{1}P^{[1]}_{1})&=&d\sigma^{L+V}(^{1}P^{[1]}_{1})+d\sigma^{S}(^{1}P^{[1]}_{1})+d\sigma^{H}(^{1}P^{[1]}_{1}) \label{eqn:tpss} \\
&=&(df^{L+V+S_1}_{^1P_1^{[1]}}+df^H_{^1P_1^{[1]}}+df^{S_2}_{^1P_1^{[1]}})\langle O^{^1P_1^{[1]}}(^1P_1^{[1]})\rangle, \NO
\eea
where $\sigma^{L+V}$ represents the summation of LO and virtual correction contributions,
and $\sigma^{S}$ and $\sigma^{H}$ are real corrections from the gluon-soft and -hard region, respectively.
The boundary of the two regions is $E_{g}=\frac{\sqrt{s}}{2} \delta_s$,
where $E_g$ is the energy of the soft gluon and $\delta_s$ is an arbitrary positive number small enough to provide the soft approximation with sufficient accuracy.
Moreover, the terms in the gluon-soft region consist of two parts,
the first of which comes from the square of the diagrams in which the soft gluon is attached to an external charm quark (antiquark) line (labeled $S_2$),
a typical one of which is displayed in Fig.~\ref{fig:gc}.
The other part comes from the rest of the amplitude squared (labeled $S_1$),
including the square of the diagrams where the soft gluon is attached to an external line other than the heavy quark line,
a typical one of which is displayed in Fig.~\ref{fig:gg},
and the interference terms of the diagrams displayed in Fig.~\ref{fig:gc} and Fig.~\ref{fig:gg}.
$f^{L+V+S_1}_{^1P_1^{[1]}}$ denotes the summation of $f^{L+V}_{^1P_1^{[1]}}$ and $f^{S_1}_{^1P_1^{[1]}}$, and is divergence free.
The only divergent term left in $d\sigma(^{1}P^{[1]}_{1})$ comes from $df^{S_2}_{^1P_1^{[1]}}$ which will subtract the divergence in $\langle O^{^1P_1^{[1]}}(^{1}S_{0}^{[8]})\rangle$.
Neglecting the finite terms proportional to the size of the small region, $d\sigma^{S_2}(^1P_1^{[1]})$ can be expressed as
\be
d\sigma^{S_2}(^{1}P^{\left[1\right]}_{1})=-\frac{\alpha_s}{3\pi m_{c}^{2}}u_{\epsilon}^{s}\frac{N_{c}^{2}-1}{N_{c}^{2}}
d f_{^{1}S_{0}^{[8]}}\langle O^{^1P_1^{[1]}}(^{1}P_{1}^{[1]})\rangle , \label{eqn:scs}
\ee
where
\be
u_{\epsilon}^{s}=\frac{1}{\epsilon_{IR}}+\frac{E}{p}ln(\frac{E+p}{E-p})+ln(\frac{4\pi\mu^{2}}{s\delta_{s}^{2}})-\gamma_{E}-\frac{1}{3} \label{eqn:us},
\ee
where $E$ and $p$ are energy and absolute value of the momentum of $h_c$.

Matching Eq.(\ref{eqn:fac}) and  Eq.(\ref{eqn:tpss}), at the same time, employing Eq.(\ref{eqn:LDM}) and  Eq.(\ref{eqn:scs}),
and constrain our calculation up to NLO in $\alpha_s$,
we obtain the expression of the short-distance coefficient for $^1P_1^{[1]}$ state as
\be
df_{^1P_1^{[1]}}^{NLO}=df^{L+V+S_1}_{^1P_1^{[1]}}+df^{H}_{^1P_1^{[1]}}-\frac{\alpha_s}{3\pi m_{c}^{2}}\frac{N_{c}^{2}-1}{N_{c}^{2}}u_{\epsilon}d f_{^1S_0^{[8]}}^{LO},
\label{eqn:soft}
\ee
where
\be
u_{\epsilon}=u_{\epsilon}^{s}-u_{\epsilon}^{c}=\frac{E}{p}ln(\frac{E+p}{E-p})+ln(\frac{\mu_{\Lambda}^{2}}{s\delta_{s}^{2}}) .
\ee

All the three terms on the right-hand side of Eq.(\ref{eqn:soft}) are finite.
Now, all the short-distance coefficients are IR divergence free, as a result, the cross section for $h_c$ hadroproduction is well defined.

Substituting Eq.(\ref{eqn:soft}) into Eq.(\ref{eqn:s}), we obtain the complete expression of the cross section for $h_c$ production,
\bea
d\sigma^{NLO}(h_{c})=\langle O^{h_c}(^1P_1^{[1]})\rangle(df^{L+V+S_1}_{^1P_1^{[1]}}+df^{H}_{^1P_1^{[1]}})
-\frac{\alpha_s}{3\pi m^{2}_{c}}\frac{N_{c}^{2}-1}{N_{c}^2}u_{\epsilon}\langle O^{h_c}(^1P_1^{[1]})\rangle df_{^1S_0^{[8]}}^{LO}
+\langle O^{h_c}(^1S_0^{[8]})\rangle df_{^1S_0^{[8]}}^{NLO} .
\label{eqn:com}
\eea

\section{numerical results}

To calculate $\sigma(^1S_0^{[8]})$ and $\sigma(^1P_1^{[1]})$,
we apply our Feynman Diagram Calculation package (FDC)~\cite{Wang:2004du} to generate all the needed FORTRAN source.

From the heavy quark spin symmetry of the NRQCD Lagrangian, it is obvious that 
the LDME $\langle O^{h_{c}}(^1S_0^{[8]})\rangle$ for the intermediate state $c\bar{c}(^{1}S^{\left[8\right]}_{0})$
evolving into $h_c$ is exactly the same as that for the intermediate state $c\bar{c}(^3S_1^{[8]})$ evolving into $\chi_{c1}$ at LO in $v^2$.
It gives
\be
\langle O^{h_{c}}(^1S_0^{[8]})\rangle\approx\langle O^{\chi_{c1}}(^3S_1^{[8]})\rangle. 
\ee

Before we present the numerical results, there is some comments on the obtaining of the CO LDME.
Focusing on the last two terms on the right-hand side of Eq.(\ref{eqn:com}),
one can notice that, if $\mu_\Lambda$ varies its value,
in order to fit the cross section $d\sigma^{NLO}(h_{c})$ to the experiment data,
the LDME in the last term should change accordingly,
which is to say, the dependence of $\mu_\Lambda$ is partly absorbed into the CO LDME.
If we proceed our calculation to infinite order in $\alpha_s$, the $\mu_\Lambda$ dependence can be totally absorbed into the CO LDME.
As a result, this scale actually can be any positive value holding the convergence of $\alpha_s$ expansion.
If our results dramatically depend on $\mu_\Lambda$, the dropped terms in higher orders must contribute significantly,
and the calculation up to this order does not reach a sufficient accuracy.
Up to NLO, the condition of $\mu_\Lambda$ independence requires
\be
\frac{\alpha_{s}}{3\pi m_{c}^{2}}\frac{N_{c}^{2}-1}{N_{c}^{2}}d f^{LO}_{^1S_0^{[8]}}\propto df^{NLO}_{^1S^{[8]}_{0}} , \label{eqn:cond}
\ee
as well as that the proportional ratio should be universal for all the processes.
Actually, in most of the cases, this condition can not be satisfied.

To show the problem of $\mu_{\Lambda}$ dependence, at the same time,
to determine whether $\alpha_s$ expansion up to NLO have got an sufficient accuracy,
one should carry out the calculation at different values of $\mu_\Lambda$.
The fitting should also be carried out at each perticular value of this scale.

In the numerical calculation, we have the following common choices as
$|R_{h_c}'(0)|^{2}=0.075\gev^5$~\cite{Eichten:1995ch} for both LO and NLO calculation, and $m_{c}=1.5\gev$.
The soft cutoff $\delta_s$ independence is checked in the calculation and $\delta_s=0.001$ is used.
Since the energy scale of most of the phase space region exceeds b-quark mass, $\Lambda_{QCD}|_{nf=5}=0.226\gev$ is used.
We employ CTEQ6M~\cite{Pumplin:2002vw} as PDF and two-loop $\alpha_s$ running for up-to-NLO calculation,
and CTEQ6L1~\cite{Pumplin:2002vw} and one-loop $\alpha_s$ running for LO.
The renormalization and factorization scales are chosen as $\mu_R=\mu_F=\sqrt{4m_c^2+p_t^2}$.

To obtain the CO LDME for $\chi_c$, we carry out a global fit,
employing experimental data in Ref.~\cite{Aaij:2011jh, LHCb:2012af, ATLAS:2014ala, Abulencia:2007bra, Aaij:2013dja, Chatrchyan:2012ub},
excluding $p_t<7\gev$ points.
The detail of the fit is presented in Ref.~\cite{Jia:2014jfa},
here, we only give the values of the CO LDMEs.
For NLO calculation, they are
\bea
&&\langle O^{h_c}(^1S_0^{[8]})\rangle_{m_c}=(7.23\pm 0.27)\times 10^{-3}\gev^3 \NO \\
&&\langle O^{h_c}(^1S_0^{[8]})\rangle_{m_c/2}=(5.49\pm 0.27)\times 10^{-3}\gev^3 \label{eqn:ldme} \\
&&\langle O^{h_c}(^1S_0^{[8]})\rangle_{\Lambda_{QCD}}=(2.52\pm 0.24)\times 10^{-3}\gev^3 \NO
\eea
where the values of $\mu_{\Lambda}$ are labeled at the foot of LDMEs.
For all the three choices of $\mu_\Lambda$, as is seen in Ref.~\cite{Jia:2014jfa},
theoretical predictions agree with the experimental data equally well.
One may notice that, here, we choose a very small scale, $\Lambda_{QCD}$.
It is not to say that this scale makes sense.
The reason is, as the readers will see afterwards, for one of the conditions we consider in this paper,
only when this scale is very small, the cross section turns out to be physical (positive).
If $\mu_\Lambda$ is larger than or of the same magnitude as $m_c/2$,
we get nonphysical results (negative cross sections).
For LO calculation, we give a band,
the upper and lower bound of which correspond to $\langle O^{h_c}(^1S_0^{[8]})\rangle=0.00013\gev^3$ and $\langle O^{h_c}(^1S_0^{[8]})\rangle=0.00188\gev^3$,
respectively, as suggested in Ref.~\cite{Jia:2014jfa}.

Employing these LDMEs, $h_c$ production rate for three experimental conditions are presented in the following.

\begin{figure}
\center{
\includegraphics*[scale=0.4]{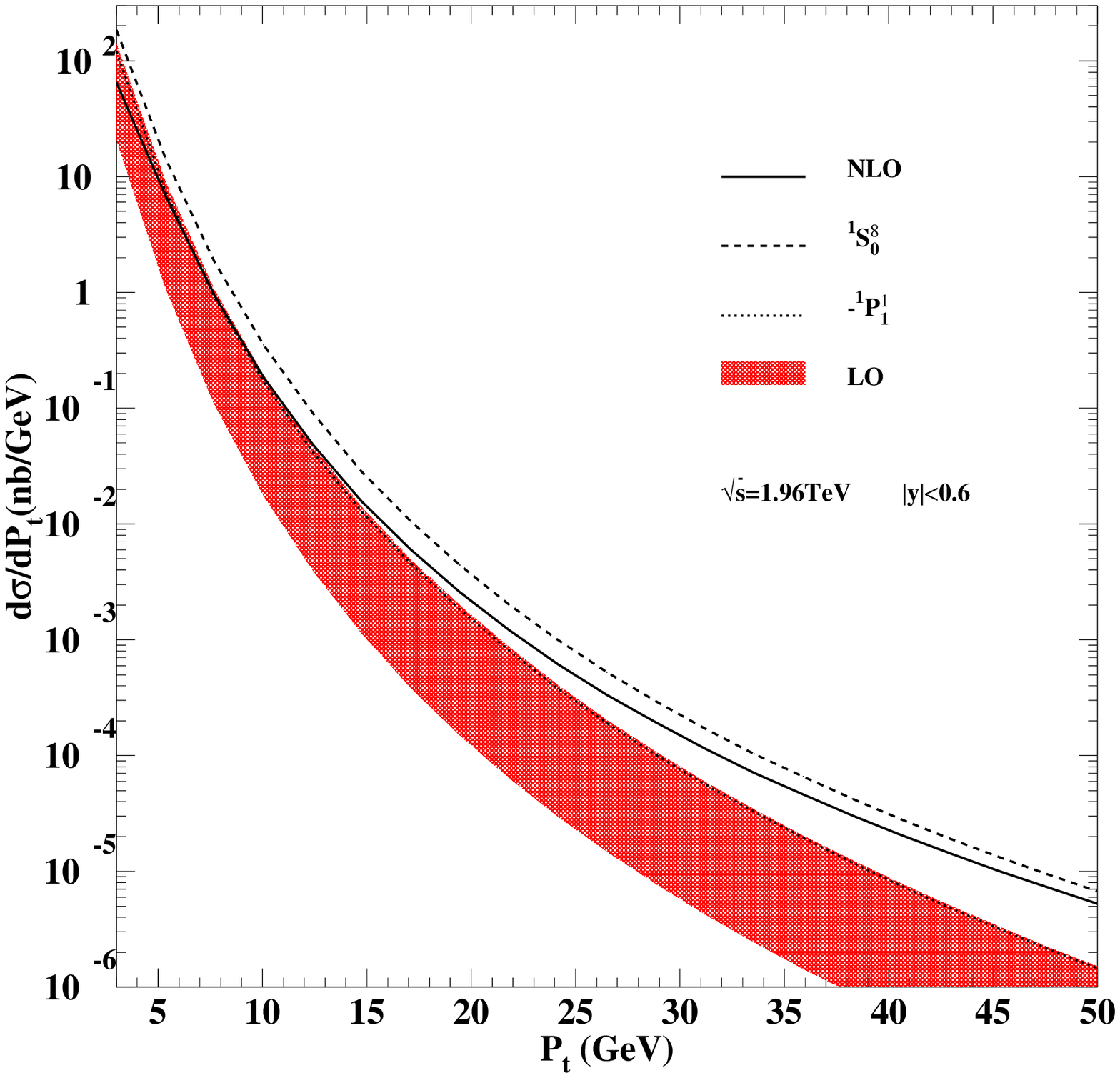}\\% Here is how to import EPS art
\includegraphics*[scale=0.4]{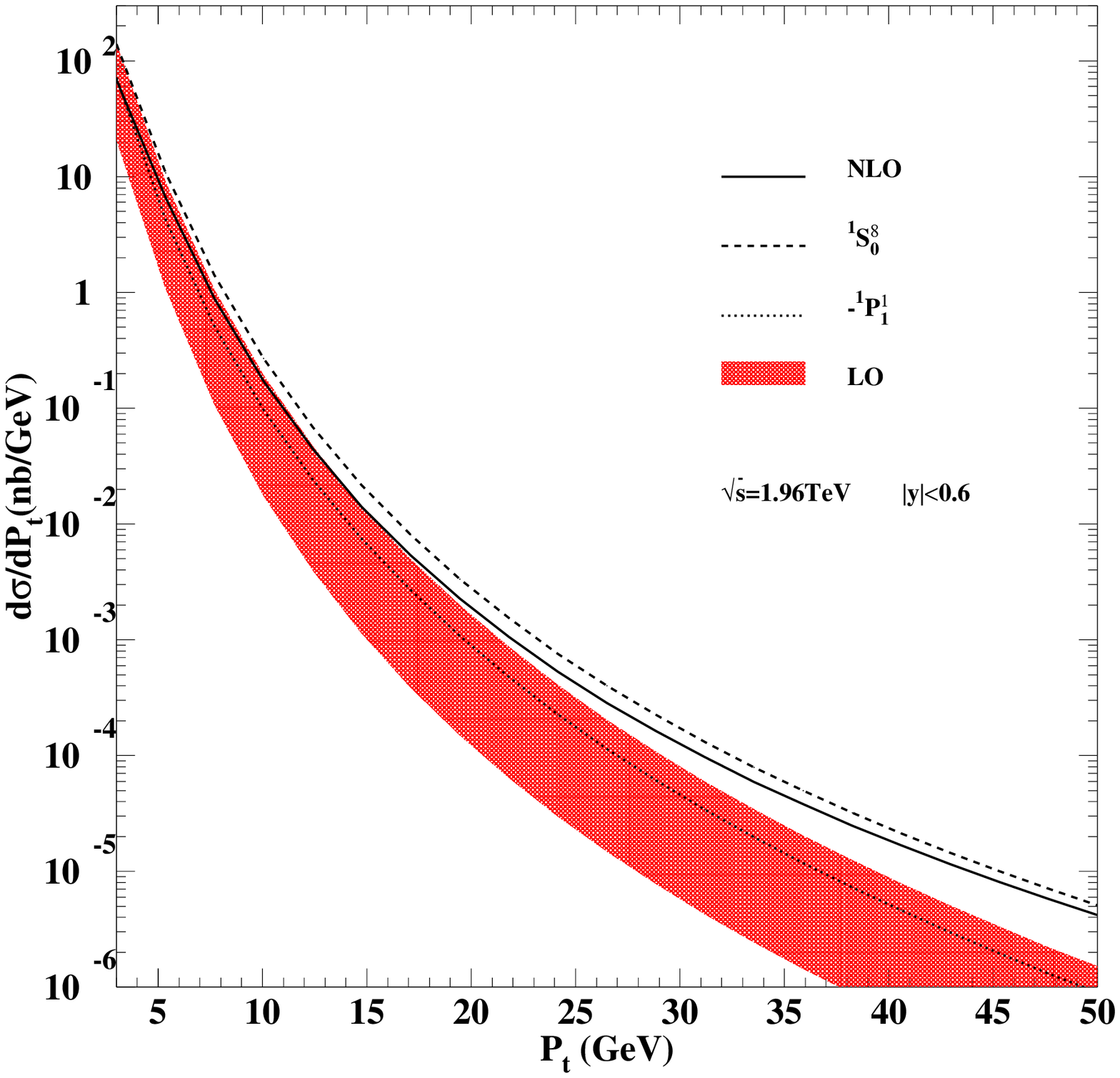}\\% Here is how to import EPS art
\includegraphics*[scale=0.4]{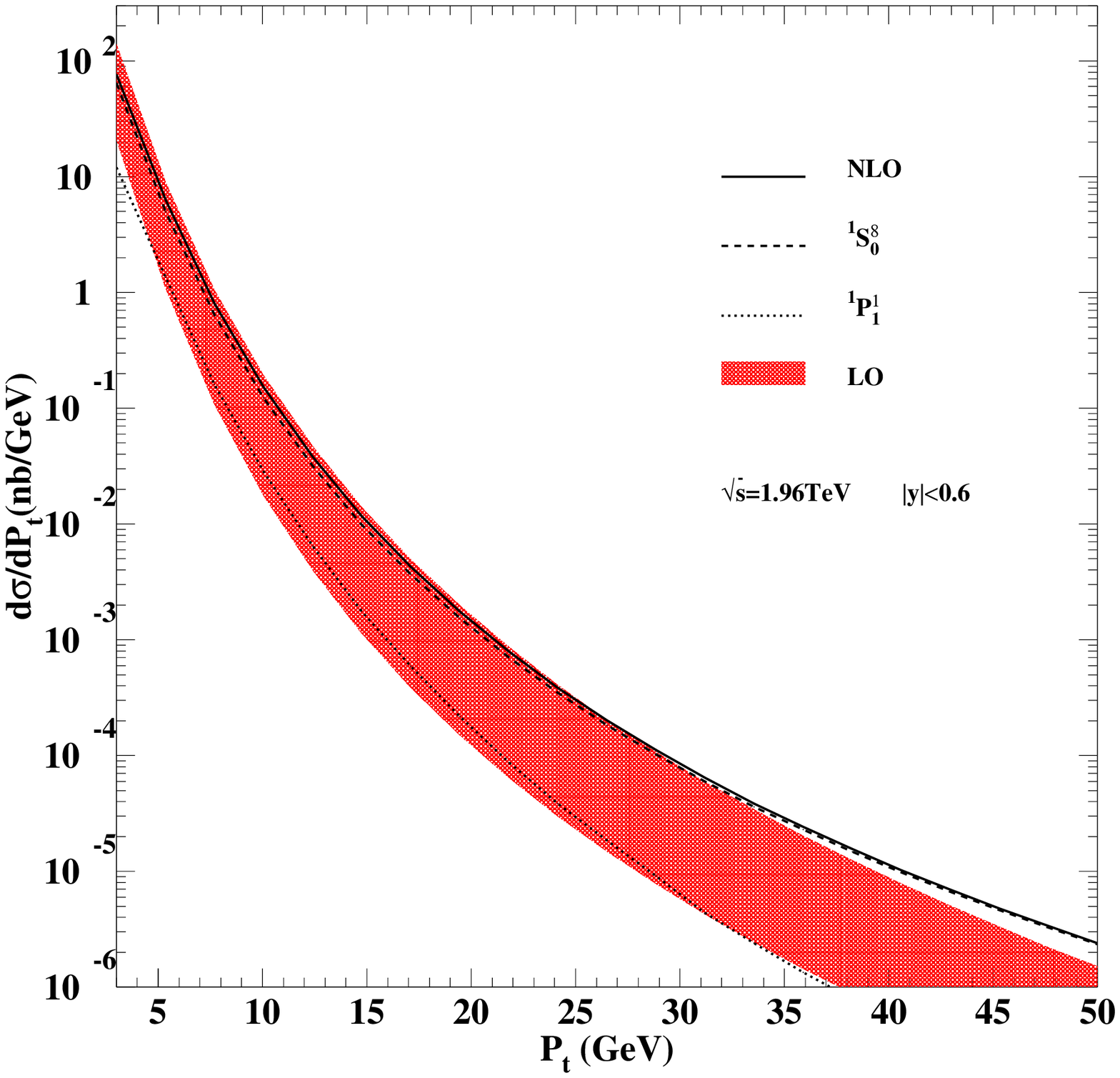}% Here is how to import EPS art
\caption {\label{fig:cdf} $h_c$ producton at the Tevatron.
The CM energy and rapidity cut are $\sqrt{s}=1960\gev$ and $|y|<0.6$, respectively.
The values of $\mu_{\Lambda}$ are $m_c$, $m_c/2$ and $\Lambda_{QCD}$ for upper, middle and lower figures, respectively.
}}
\end{figure}

\begin{figure}
\center{
\includegraphics*[scale=0.4]{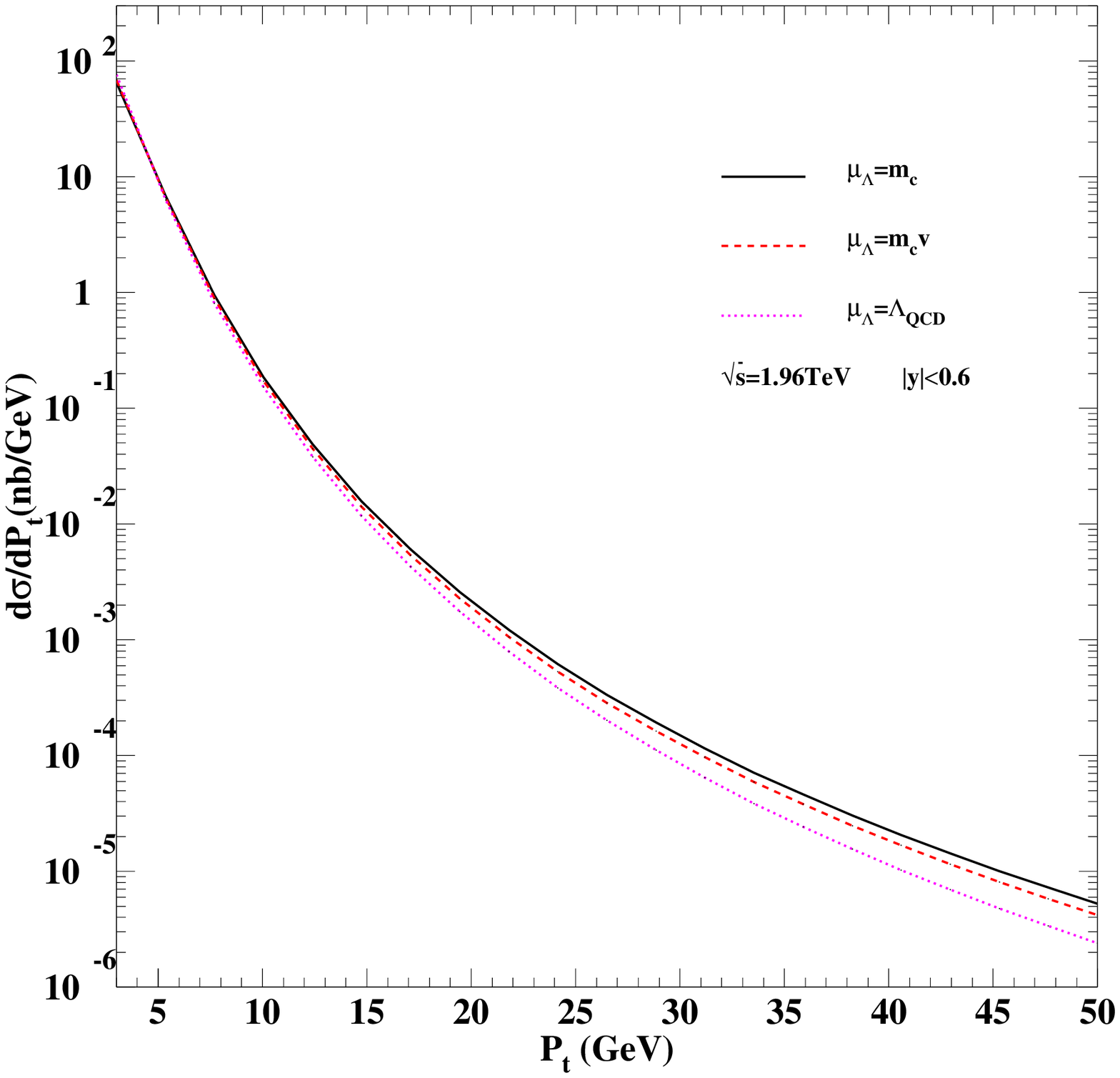}% Here is how to import EPS art
\caption {\label{fig:cdfcom}
The comparison of the results for the three different choices of NRQCD scale.
The CM energy and rapidity cut are $\sqrt{s}=1960\gev$ and $|y|<0.6$, respectively.
}}
\end{figure}

For Tevatron energy, i.e. $\sqrt{s}=1.96$TeV, and rapidity cut condition $|y|<0.6$,
denoted as experimental condition I (EC1), $h_c$ production rates are presented in Fig.~\ref{fig:cdf}.
As is shown in these figures, despite the contributions from the two channels,
$^1S_0^{[8]}$ and $^1P_1^{[1]}$, vary significantly for different choices of $\mu_\Lambda$,
the final results do not change much.
The distinction of the final results for the three scale choices is presented more explicitly in Fig.~\ref{fig:cdfcom}.
We can see that, in low $p_t$ region, different values of $\mu_\Lambda$ bring in little distinction,
while in high $p_t$ region, the three curves differ, but not much (the largest is about twice of the smallest).
Since large logarithm term $ln(m_c/p_t)$ may ruin $\alpha_s$ expansion for both too large and too small $p_t$,
besides, in small $p_t$ region, relativisitic corrections contribute a nonlinear remarkable part,
only medium $p_t$ region is considered in fixed order calculation.
Restricted to this region, our results do not depend on NRQCD scale significantly.
We can say the dependence of $\mu_\Lambda$ has been absorbed into the CO LDME.

\begin{figure}
\center{
\includegraphics*[scale=0.4]{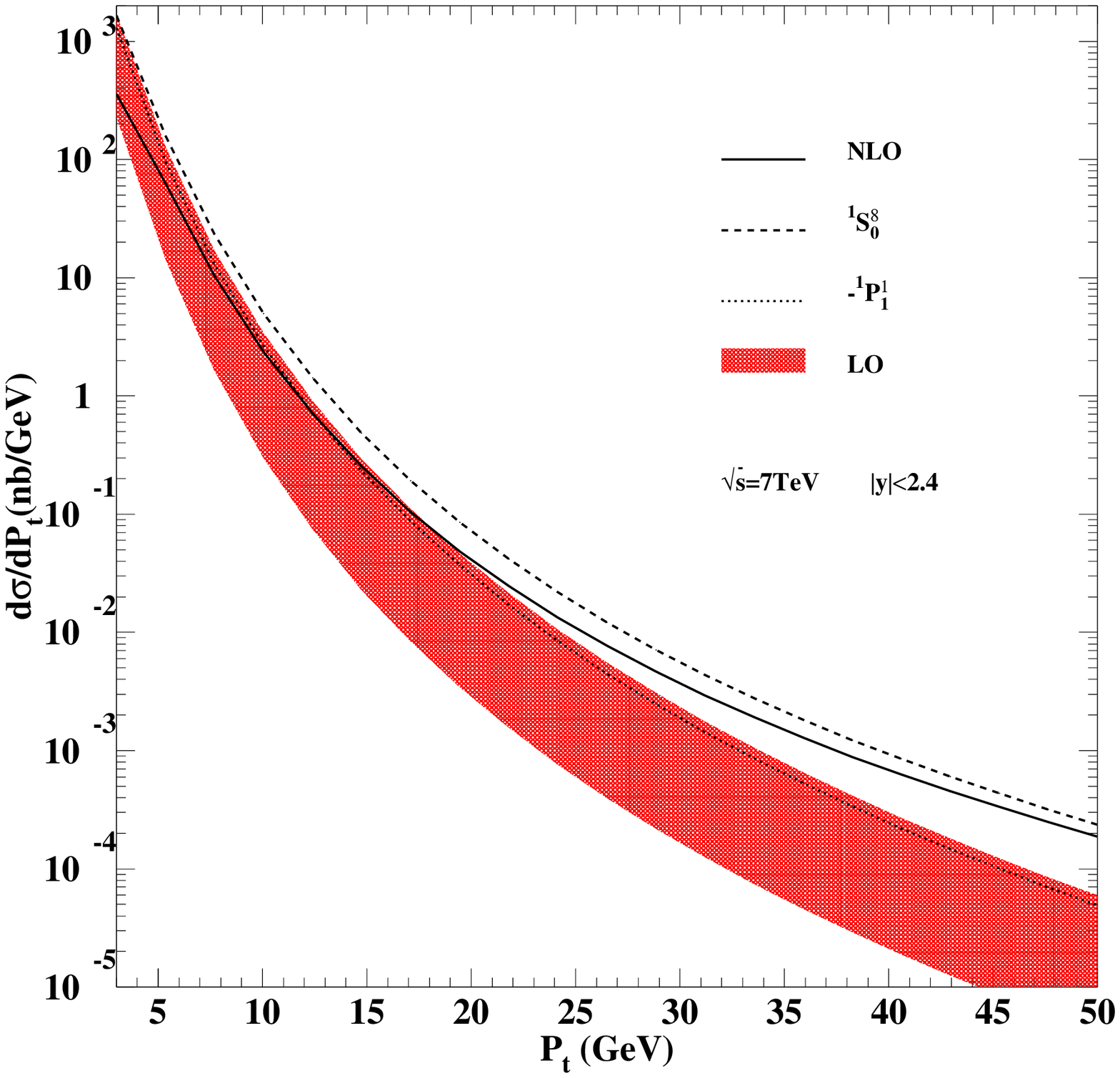}\\% Here is how to import EPS art
\includegraphics*[scale=0.4]{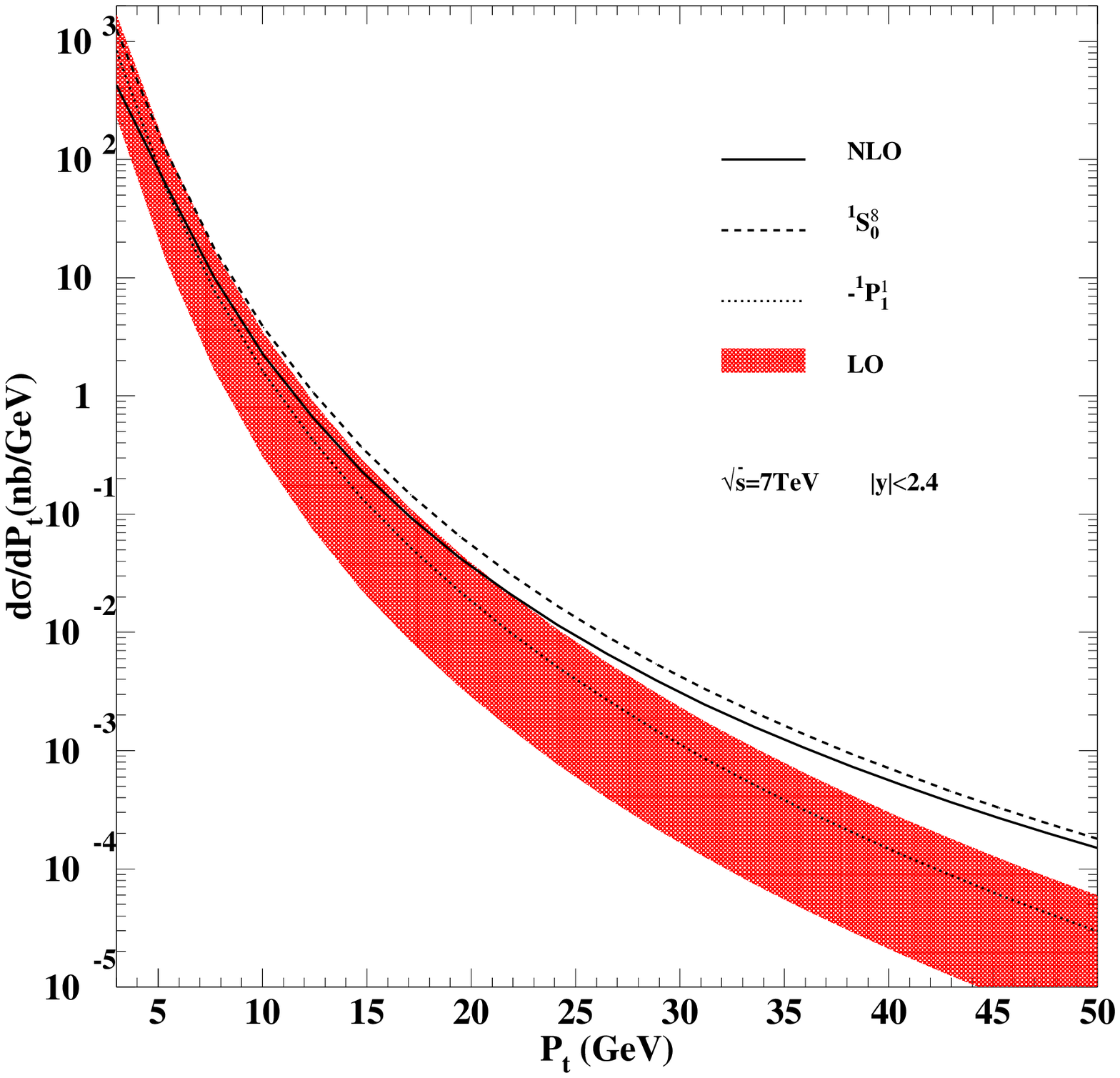}\\% Here is how to import EPS art
\includegraphics*[scale=0.4]{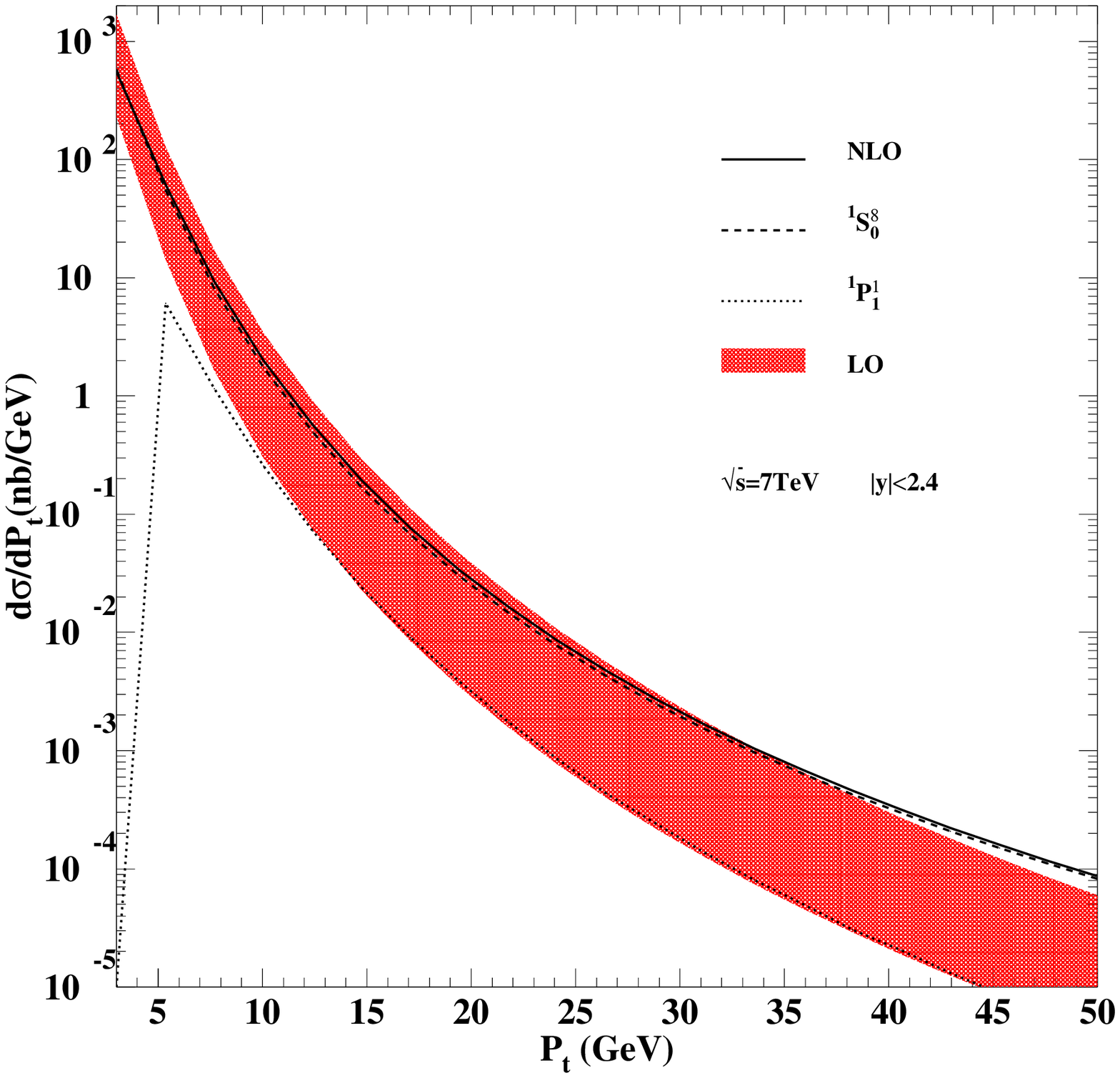}% Here is how to import EPS art
\caption {\label{fig:atlas} $h_c$ producton at the LHC.
The CM energy and rapidity cut are $\sqrt{s}=7000\gev$ and $|y|<2.4$, respectively.
The values of $\mu_{\Lambda}$ are $m_c$, $m_c/2$ and $\Lambda_{QCD}$ for upper, middle and lower figures, respectively.
}}
\end{figure}

\begin{figure}
\center{
\includegraphics*[scale=0.4]{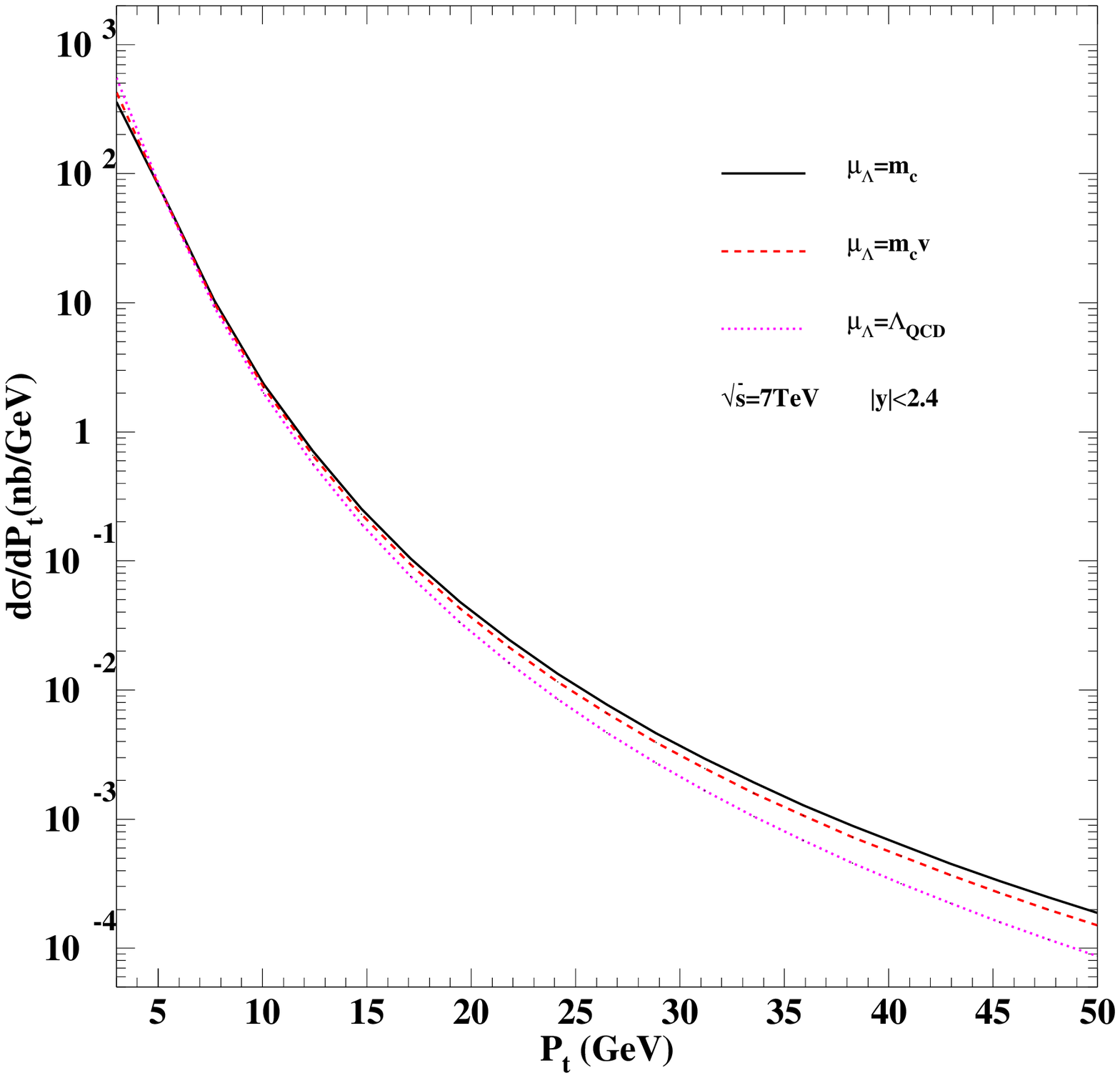}% Here is how to import EPS art
\caption {\label{fig:atlascom}
The comparison of the results for the three different choices of NRQCD scale.
The CM energy and rapidity cut are $\sqrt{s}=7000\gev$ and $|y|<2.4$, respectively.
}}
\end{figure}

For LHC energy $\sqrt{s}=7$TeV, while rapidity range being $|y|<2.4$,
denoted as experimental condition II (EC2), we got similar conclusion;
NRQCD scale dependence is not so severe to ruin the accuracy of prediction,
as is shown in Fig.~\ref{fig:atlas} and Fig.~\ref{fig:atlascom}.

\begin{figure}
\center{
\includegraphics*[scale=0.4]{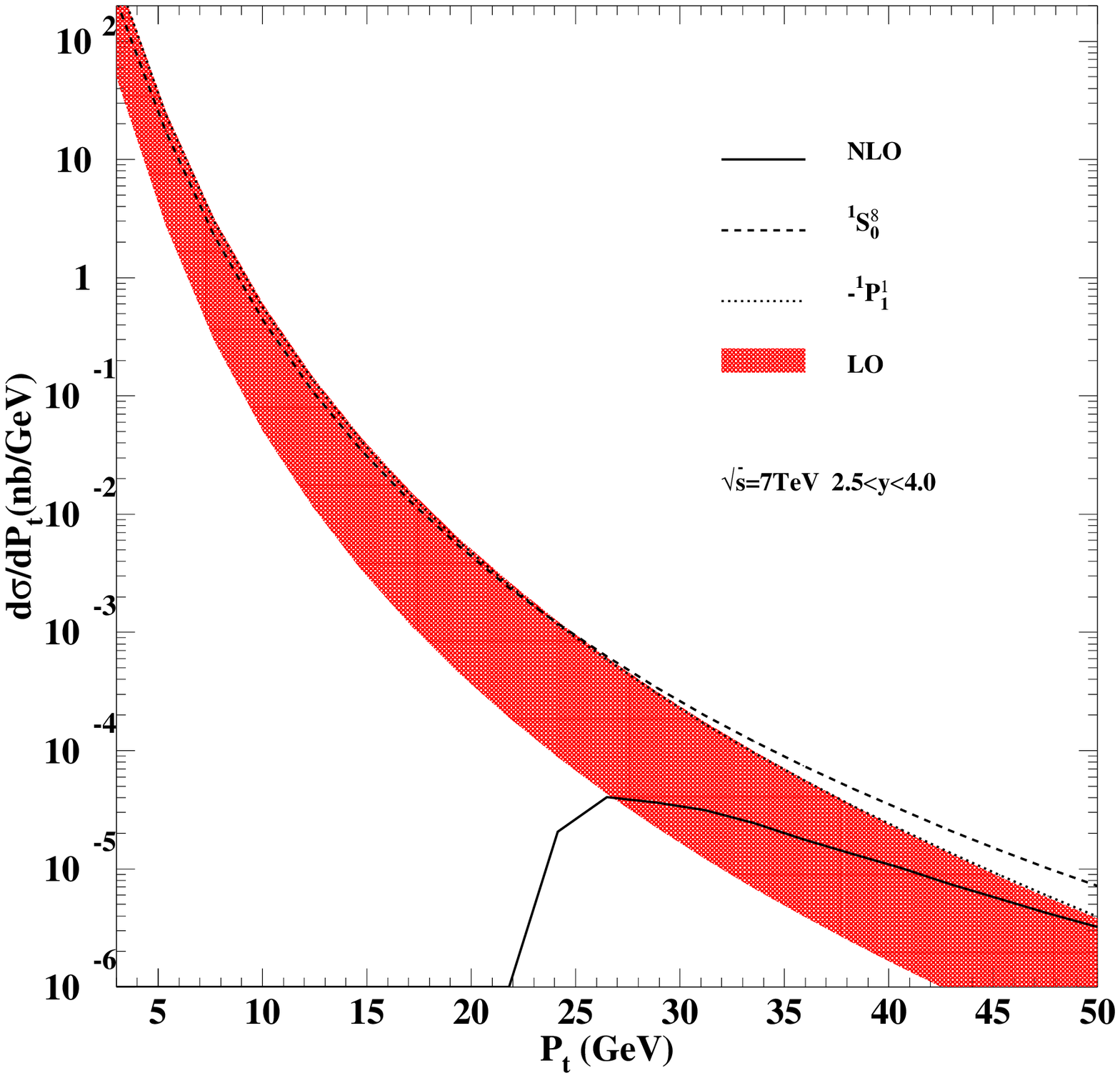}\\% Here is how to import EPS art
\includegraphics*[scale=0.4]{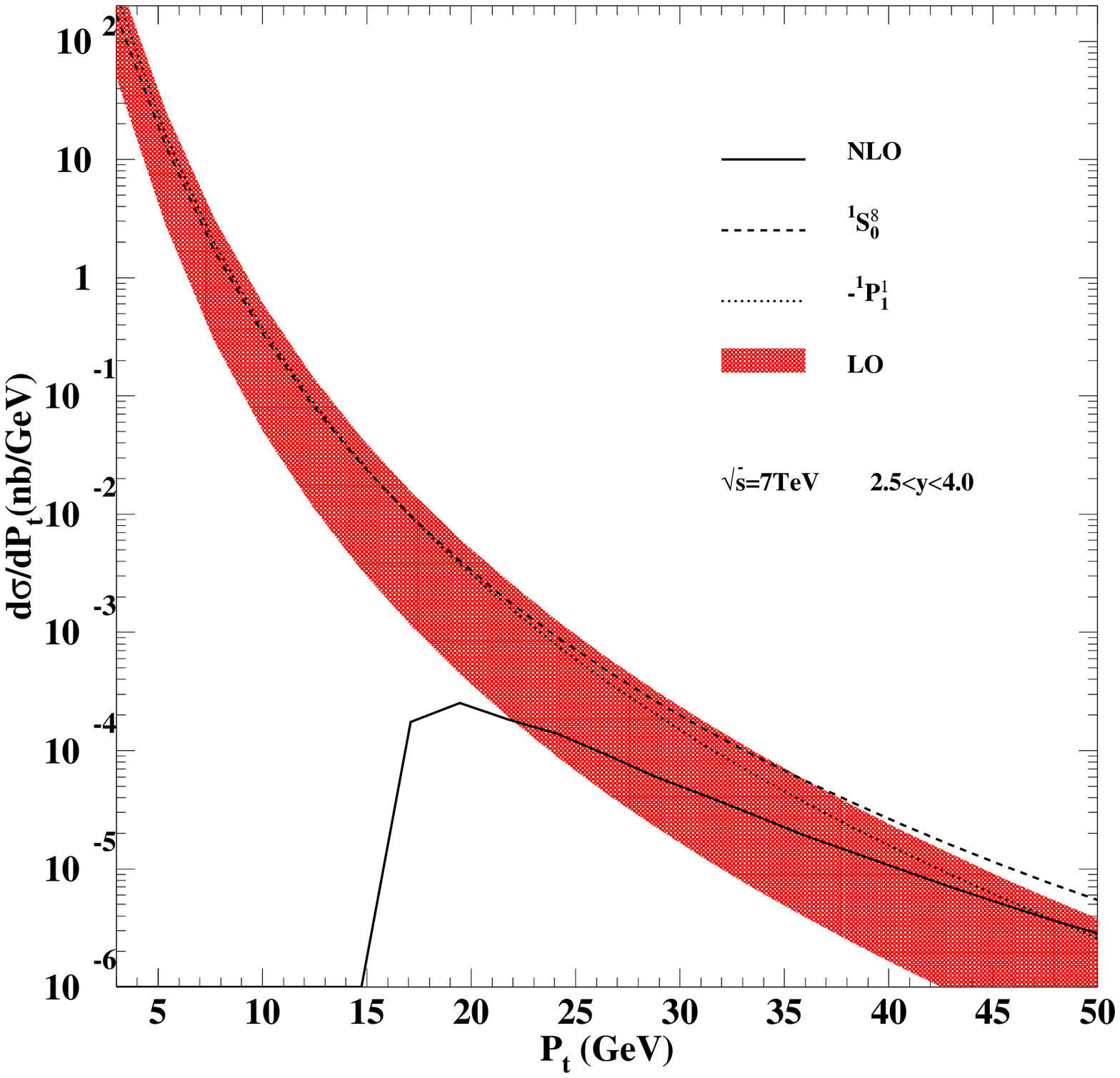}\\% Here is how to import EPS art
\includegraphics*[scale=0.4]{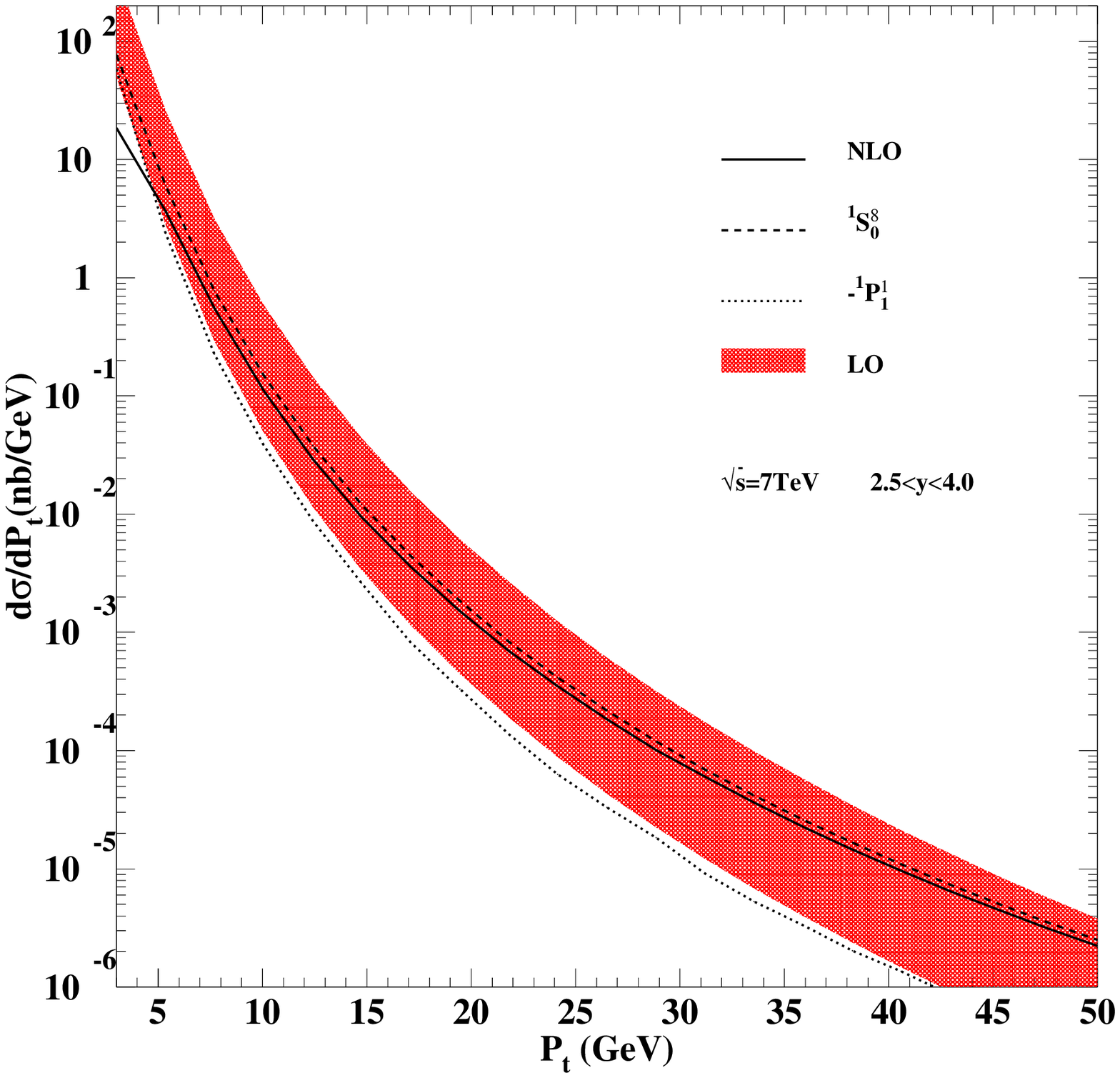}% Here is how to import EPS art
\caption {\label{fig:alice} $h_c$ producton at the LHC.
The CM energy and rapidity cut are $\sqrt{s}=7000\gev$ and $2.5<y<4$, respectively.
The values of $\mu_{\Lambda}$ are $m_c$, $m_c/2$ and $\Lambda_{QCD}$ for upper, middle and lower figures, respectively.
}}
\end{figure}

\begin{figure}
\center{
\includegraphics*[scale=0.4]{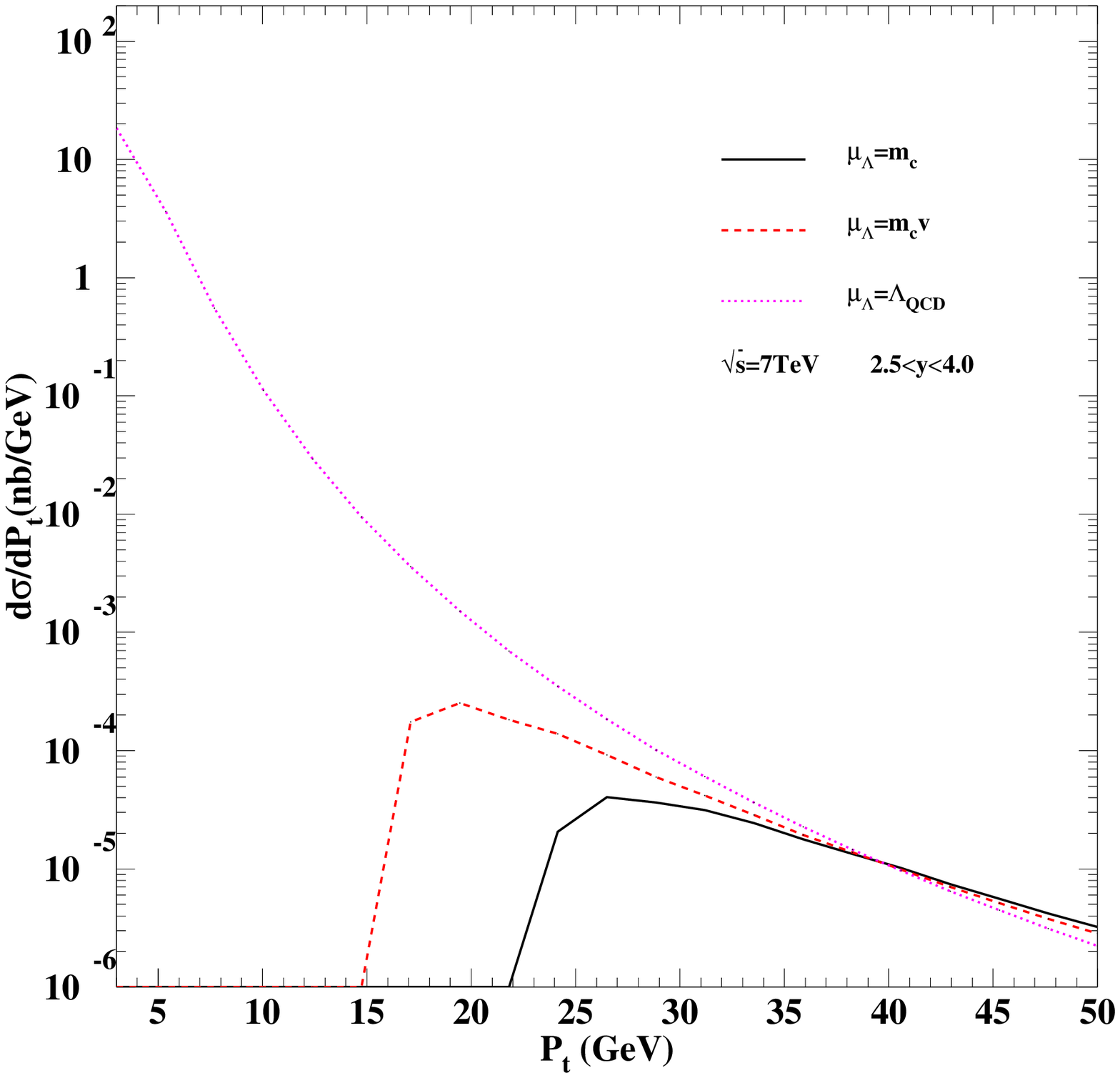}% Here is how to import EPS art
\caption {\label{fig:alicecom}
The comparison of the results for the three different choices of NRQCD scale.
The CM energy and rapidity cut are $\sqrt{s}=7000\gev$ and $2.5<y<4$, respectively.
}}
\end{figure}

However, at the same LHC energy, for $2.5<y<4.0$,
denoted as experimental condition III (EC3), when $\mu_{\Lambda}=m_c$ and $m_c/2$,
we got unphysical results, i.e. negative cross sections (as is shown in Fig.\ref{fig:alice}).
The dependence of NRQCD scale is so severe that we cannot make a definite conclusion for this experimental condition.
Only for $\mu_{\Lambda}=\Lambda_{QCD}$, the cross section turns out to be positive through out the whole range of $p_t$.
The comparison of the results for the three different choices of $\mu_\Lambda$ is also presented, as is shown in Fig.~\ref{fig:alicecom}.
We notice that, for a proper determined NRQCD scale $\mu_\Lambda=m_c$ (default choice),
i.e. it is comparable to other scales in this calculation and in perturbative region,
even for medium $p_t$, we come across nonphysical results,
which might be caused by a new scale, the energy of $h_c$ ($E(h_c)$), brought in by large $y$: $E(h_c)\approx m_te^y/2$.
Before resumming large log terms brought in by this scale, one can not obtain reliable results.

\section{analysis on $\mu_\Lambda$ dependence}

\begin{figure}
\center{
\includegraphics*[scale=0.5]{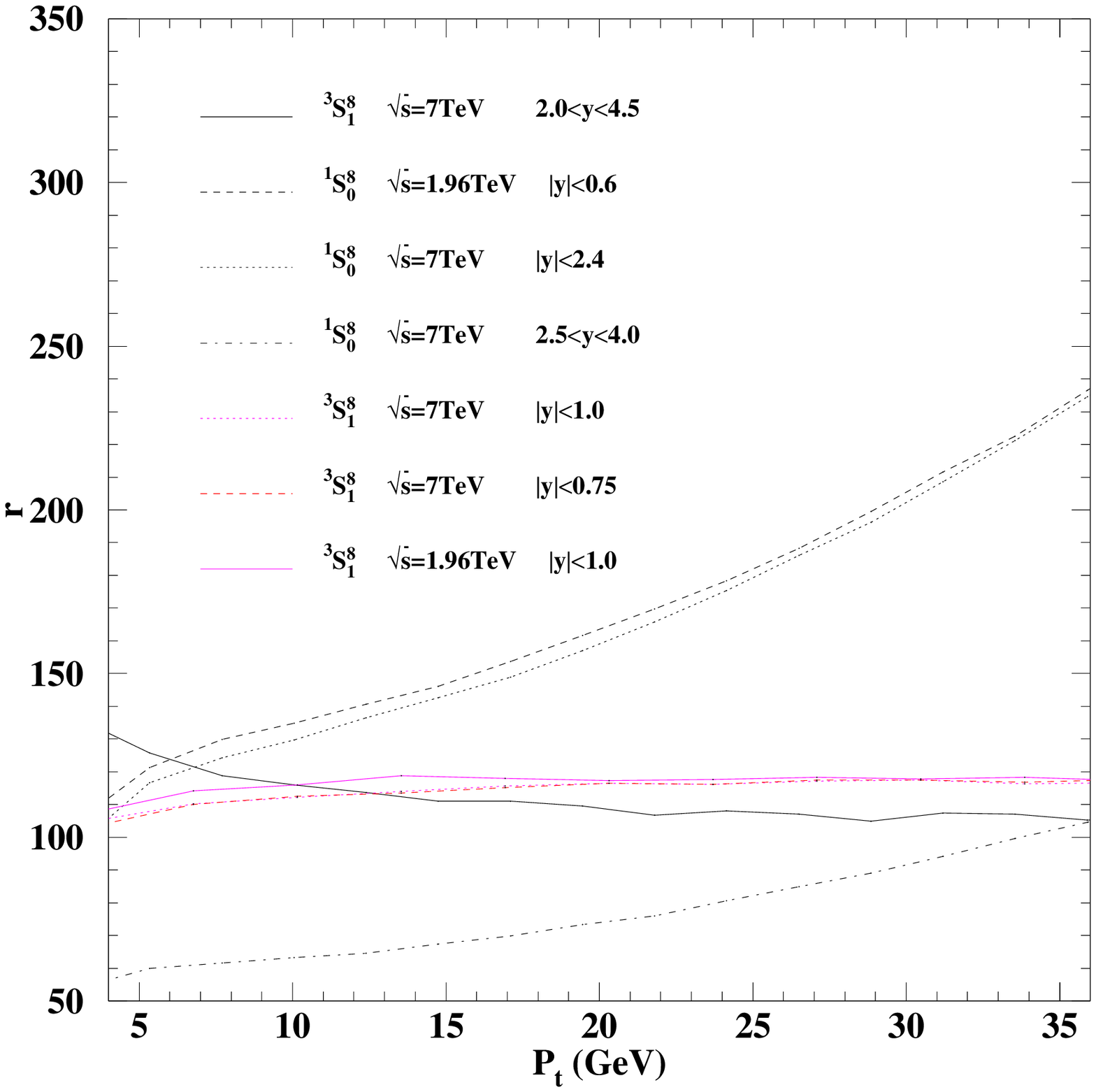}% Here is how to import EPS art
\caption {\label{fig:ratio}
The comparison of the value of $r$ for the three different experimental conditions for $h_c$, as well as $\chi_c$, as a function of $p_t$.
The CM energy for LHC and Tevatron experiment are $\sqrt{s}=7000\gev$ and $\sqrt{s}=1960\gev$, respectively.
}}
\end{figure}

In order to investigate the origin of the discrepency of the three curves in Fig.\ref{fig:alicecom},
we define the proportional ratio for Eq.(\ref{eqn:cond}) as
\be
r=\frac{df^{NLO}_{n}}{dp_t}/(\frac{\alpha_{s}}{3\pi}\frac{N_{c}^{2}-1}{N_{c}^{2}}\frac{d f^{LO}_{n}}{dp_t}), \label{eqn:defr}
\ee
where $n$ is either $^1S_0^{[8]}$ or $^3S_1^{[8]}$.
We calculate this parameter for $h_c$ productions for the three experimental conditions,
as well as that for $\chi_c$ production for the four experimental conditions we used in the fit.
Fig.~\ref{fig:ratio} compares the values of $r$ for the seven conditions.
In the case of $\chi_c$ production, the dependence of $r$ on $p_t$ is flat,
which explains the fact that, the experiment on $\chi_c$ can be fitted well for all the three choices of $\mu_\Lambda$.
At small and medium $p_t$, the curve for $\chi_c$ is close to that for $h_c$ in EC1 and EC2.
This results in slight $\mu_{\Lambda}$ dependence in the two conditions, just as shown in Fig.~\ref{fig:cdfcom} and ~\ref{fig:atlascom}.
However, for EC3, the value of $r$ is quite (about two and a half time) below that for $\chi_c$,
which causes the scale dependence problem, as is shown in Fig.~\ref{fig:alicecom}.
In high $p_t$ region, the curves for EC1 and EC2 are quite above that for $\chi_c$,
the corresponding effect is that, in this $p_t$ region,
the dependence of the production rates of $h_c$ on NRQCD scale is more remarkable than in low $p_t$ region.
By contrast, this problem for EC3 becomes milder in high $p_t$ region.

\section{summary and outlook}

The NRQCD factorization framework has gained its reputation for its success in many aspects since its discovery,
especially for P-wave quarkonium productions and decays,
where CSM fails for IR singularities, and NRQCD deals with the problem.
However, at the same time, it introduces another scale, the NRQCD scale $\mu_\Lambda$.
Before our work, most believed that $\mu_\Lambda$ dependence can always be absorbed into CO LDMEs,
which is, however, not true for some cases.
This paper studies the NRQCD scale dependence problem for the first time,
and provide a brief and phenomelogical analysis on this problem.
We can see in Section IV that, the ratio $r$ is a crucial quantity to investigate $\mu_\Lambda$ dependence at NLO.
When $r$ is far from a constant, just as is shown in Fig.\ref{fig:ratio},
one fixed value of the CO LDME cannot give reasonable predictions in all the phase space regions for all the experimental conditions.
One possible solution is that one resum the terms giving rise to $\mu_\Lambda$ dependence.
The requirement of $r$ to be a constant is also a necessity for the perturbative calculation up to NLO to reach a sufficient precision.
As a result, to draw a definite conclusion,
one should also calculate this value to determine whether its calculation up to NLO can provide trustable results.

In addition, this paper presents the QCD NLO theoretical predictions of the $h_c$ production at the Tevatron and the LHC for three experimental conditions.
LO curves are also presented, using LDMEs provided in Ref.~\cite{Jia:2014jfa}.
Using the CO LDMEs for different values of NRQCD scale,
we study the NRQCD scale ($\mu_\Lambda$) dependence of $h_c$ hadroproduction rate for the three experimental conditions.
In medium $p_t$ region, where perturbative calculation is available,
for EC1 and EC2, the final results depend on $\mu_\Lambda$ slightly,
as a result, theoretical predictions up to this order are reliable.
The production rates are physical for all the choices of $\mu_\Lambda$ considered in this paper for these two conditions.
By contrast, for EC3, we obtain a negative production rate in low (and medium) $p_t$ region while setting $\mu_\Lambda$ as $m_c$ (and $m_c/2$).
In EC3, the theoretical prediction remarkably depend on $\mu_\Lambda$ in medium $p_t$ region,
as a result, theoretical prediction for this experimental condition fails at NLO.
Our calculations for LO results agree with Ref.~\cite{Qiao:2009zg} when employing the same choices of parameters,
yet cannot accord with Ref.~\cite{Sridhar:1996vd, Sridhar:2008sc} despite having tried all the possible choices of the parameters.

This work is supported by the National Natural Science Foundation of
China (Nos.~11405268, 10979056 and 10935012), in part by DFG and NSFC (CRC 110) and CAS under Project No. INFO-115-B01.

%\bibliography{paper}% Produces the bibliography via BibTeX.

\end{document}